\documentclass[aps,prl,reprint,floatfix,a4paper,raggedbottom,nobalancelastpage]{revtex4-1}  
\pdfoutput=1
\usepackage{cmap}
\usepackage[T1]{fontenc}
\usepackage{mathptmx}
\usepackage{times}
\usepackage{fixmath}

\DeclareSymbolFont{UPM}{U}{eur}{m}{n}  
\DeclareMathSymbol{\partial}{0}{UPM}{"40}

\usepackage{helvet}

\usepackage[pdftex, bookmarks, pdffitwindow=false, pdfstartview=FitH, pdfdisplaydoctitle, colorlinks, plainpages=false, pdftitle={Numerical evidence for nucleated self-assembly of DNA brick structures},pdfauthor={Aleks Reinhardt, Daan Frenkel}, pdfpagelabels, hypertexnames, citecolor={blue},linkcolor={red}, urlcolor={violet}, pdflang={en}, hyperfootnotes=false, breaklinks]{hyperref}

\usepackage{graphicx}
\usepackage{amsmath}

\usepackage{geometry}
\usepackage{setspace}

\usepackage[stretch=15,shrink=15,step=1]{microtype}
\usepackage[version=3,arrows=pgf]{mhchem}
\usepackage{siunitx}

\usepackage[nodayofweek]{datetime}
\newdateformat{myDate}{\THEDAY\ \monthname[\THEMONTH] \THEYEAR}

\begin{document}

\title{Numerical evidence for nucleated self-assembly of DNA brick structures}
\author{Aleks Reinhardt}
\author{Daan Frenkel}
\email[Correspondence author. E-mail: ]{df246@cam.ac.uk}
\affiliation{Department of Chemistry, University of Cambridge, Lensfield Road, Cambridge, CB2 1EW, United Kingdom}
\date{\myDate\today}

\raggedbottom

\begin{abstract}
The observation by Ke~\textit{et~al.}~[Science \textbf{338}, 1177 (2012)] that large numbers of short, pre-designed DNA strands can assemble into three-dimensional target structures came as a great surprise, as no colloidal self-assembling system has ever achieved the same degree of complexity. That failure seemed easy to rationalise: the larger the number of distinct building blocks, the higher the expected error rate for self-assembly. The experiments of Ke~\textit{et~al.}~have disproved this argument. Here, we report Monte Carlo simulations of the self-assembly of a DNA brick cube, comprising approximately 1000 types of DNA strand, using a simple model. We model the DNA strands as lattice tetrahedra with attractive patches, the interaction strengths of which are computed using a standard thermodynamic model. We find that, within a narrow temperature window, the target structure assembles with high probability. Our simulations suggest that mis-assembly is disfavoured because of a slow nucleation step. As our model incorporates no aspect of DNA other than its binding properties, these simulations suggest that, with proper design of the building blocks, other systems, such as colloids, may also assemble into truly complex structures.
\end{abstract}



\maketitle

\raggedbottom

The development of DNA `origami'~\cite{Seeman2003, *Seeman1982, Winfree1998, Rothemund2006} has made it possible to exploit the exquisite designability of DNA hybridisation to create a range of novel, self-assembling structures that promise to have applications in virtually all aspects of nanotechnology (for a review, see Ref.~\citenum{Linko2013}).  The original version of DNA origami employed a long `scaffold' single-stranded (ss)DNA sequence and linking `staple' ssDNA molecules that serve to fold the scaffold strand into the desired shape \cite{Rothemund2006}. A variety of structures have been assembled, including simple sheets, boxes that can open and close, `smiley faces' and curved vase-like containers \cite{Torring2011}.

In 2012, Ke~\textit{et~al.}~reported a radically different approach that dispenses with the long ssDNA template \cite{Ke2012}. Their method is based on the pre-fabrication of small DNA bricks that can be linked together in a way somewhat akin to Lego bricks, but Lego bricks that fit in only one pre\-determined part of the target structure. With this approach, it proved possible to construct almost any target structure up to a given size simply by preparing a mixture of the designed DNA bricks and cooling it down. This makes structure design considerably simpler than traditional DNA origami synthesis, in which a new set of staple strands must be designed for every new shape one wishes to construct. Moreover, while traditional DNA origami takes the scaffold strand from viral DNA, no biological DNA is required in DNA brick assembly. Ke~\textit{et~al.}~demonstrated the applicability of their approach by constructing over 100 shapes from a cuboid `canvas' \cite{Ke2012}, and this modular design has also been used to construct two-dimensional structures \cite{Wei2012, Zhang2013c} and more complex building blocks \cite{Wei2013}.

It should be stressed that the observation of Ref.~\citenum{Ke2012} was very surprising. The  self-assembly of short ssDNA strands may seem intuitive at first glance, given that DNA provides for precise sequence matching to allow only the correct `bricks' to stick together, but in the self-assembly of (say) a molecular crystal, self-poisoning is a serious problem: if molecules are incorporated  incorrectly in the crystal, the target structure cannot be reached. Apparently, DNA bricks manage to avoid this issue.  This fact is even more surprising since the bricks of Ref.~\citenum{Ke2012} were made using `positive' design only, whereby the favourable interactions between putative neighbours were chosen, but no `negative' design \cite{Winfree1998}, \textit{i.e.}~without excluding possible undesired interactions. With many copies of each DNA strand in the system, the potential for incorrect assembly is significant. Indeed, templated DNA origami was developed precisely to avoid this problem \cite{Rothemund2012}.  Ke~\textit{et~al.}~suggest that in their system, seeding is slower than the subsequent growth of the desired structure, thereby minimising the tendency for incorrect assembly, but it is not immediately obvious that this should be the case \cite{Gothelf2012}.

The aim of this Letter is to explore whether a generic, and absolutely minimal, model of DNA bricks can reproduce the findings of Ke~\textit{et~al.}: if this were to be the case, this would be good news, because it would imply that similar complex structures could be made with very different building blocks, provided they had the same functionality as DNA bricks.

The basic principle of DNA self-assembly design is that the target structure has the lowest free energy, which is usually realised by maximising complementary Watson--Crick base pairings \cite{Doye2013}. However, the self-assembly kinetics are not well understood, and more specifically, we do not know what it takes to avoid kinetic traps \cite{Doye2013}. Studying the DNA brick self-assembly process in detail would allow us to gain an understanding of the factors governing the rates and yields associated with the process and might eventually assist in the formulation of optimal design rules.

As DNA brick structures comprise several thousand base pairs, all-atom simulations long enough to observe self-assembly would be prohibitively time consuming. A coarse-grained model is therefore needed, but such a model, whilst simple, should not be \emph{too} simple: it should capture the essential features of real DNA hybridisation.  While several coarse-grained models have been developed in recent years \cite{Doye2013}, most of these are still much too detailed to be usable in studying DNA brick assembly.

In deciding on the principal physical features that must be retained in a coarse-grained description suitable for assembling DNA brick structures, we first consider some aspects of the experimental system of Ref.~\citenum{Ke2012}, in which each 32-nucleotide ssDNA molecule bonds with four other molecules through a quarter of its total length (called a `domain') to form the final structure. Each double-stranded segment thus comprises 8 base pairs, which gives a dihedral angle of $\sim$\ang{90} \cite{Ke2012}. Of particular interest is the property that, if we consider the centres of mass of each ssDNA in the final structure, these form a distorted diamond lattice \cite{Ke2012}. This suggests that we can describe each molecule, when bonded, as a tetrahedron to a first approximation. Therefore, in our approach, each single-stranded molecule is modelled as a particle with four distinct, tetrahedrally arranged patches, and each of these patches has an associated DNA sequence.

We carry out our simulations on a cubic lattice with lattice parameter $a$. Particles interact if they are diagonally adjacent to each other, and the minimum distance between any two particles is $a\sqrt{3}$. Particle interactions are initially slightly repulsive ($\varepsilon_\text{init}/k_\text{B}=\SI{100}{\kelvin}$) to prevent large-scale agglomeration, but to this interaction energy we add the hybridisation free energy for the longest complementary ($5^\prime$-$3^\prime$/$3^\prime$-$5^\prime$) sequence match between the closest pair of `patches', allowing for single internal mismatches. This free energy is determined using the nearest-neighbour parameterisation of SantaLucia~Jr and co-workers \cite{SantaLucia2004}, where we take into account terminal A-T penalties, internal mismatches \cite{Allawi1997, *Allawi1998, *Allawi1998b, *Allawi1998c, *Peyret1999}, dangling ends \cite{Bommarito2000} and the temperature and salt concentration dependence \cite{Koehler2005}, but do not consider loops or bulges, which we do not expect to be important for sequences of at most 8 base pairs. We perform Monte Carlo (MC) simulations \cite{Metropolis1953} in the canonical ensemble with a periodic simulation box. We allow clusters of particles to move or rotate concurrently using the virtual move MC algorithm \cite{Whitelam2007, *Whitelam2008} to improve sampling efficiency. Particles (or clusters) are randomly translated along the lattice or rotated in one of the 24 predetermined orientations associated with a tetrahedron with 4 distinct vertices placed within a cube. To find the free energy as a function of the size of the largest correctly bonded cluster, we run umbrella sampling simulations \cite{Torrie1977, *Mezei1987} with umbrella sampling steps performed every \num{200000} MC steps \cite{Hetenyi2002}.

\begin{figure}[b!]
\centering
\includegraphics{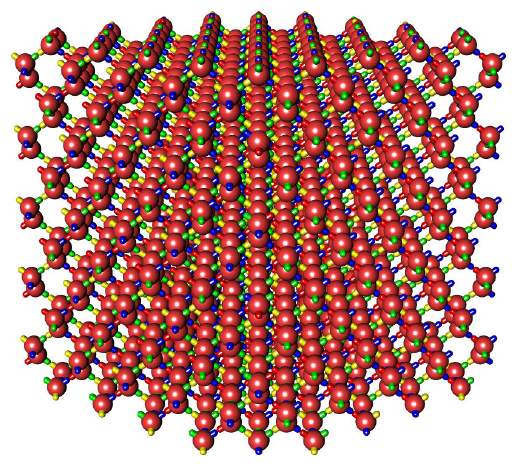}
\caption{Target structure. Each patch is colour-coded: by design, red patches bond with blue ones and green patches with yellow ones, but each patch has its own sequence.}\label{fig-bricks-snapshot-target}
\end{figure}

\begin{figure*}[tbp]
\centering
\includegraphics{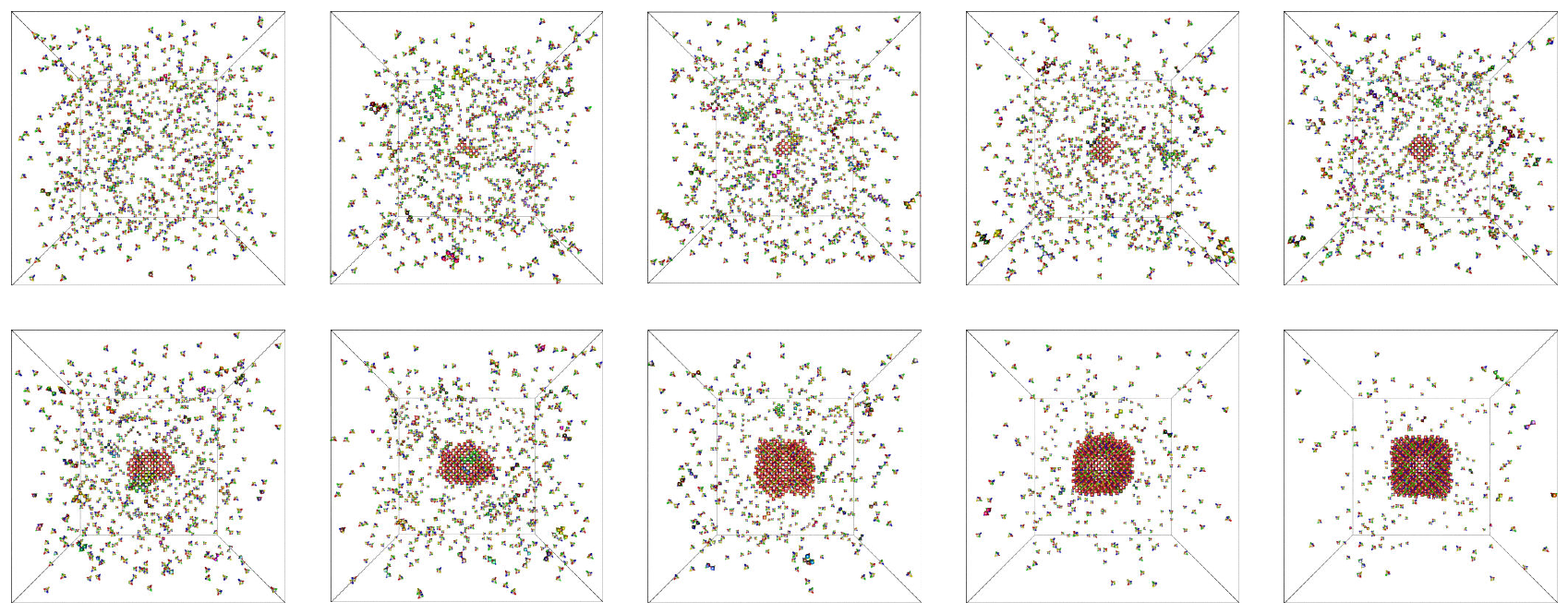}
\caption{Snapshots from a single trajectory at $T=\SI{318}{\kelvin}$, taken at $\sim$\num{2e10} step intervals and arranged in sequence. The largest correctly bonded cluster is shown in red at the centre of the simulation box; other large clusters are shown in other colours.}\label{fig-bricks-snapshots-assemblyT318}
\end{figure*}

Our target structure comprises 998 ssDNA molecules that, when correctly assembled, form a cube (Fig.~\ref{fig-bricks-snapshot-target}). Ke~\textit{et~al.}~found that randomly selected sequences that fulfil the bonding requirements have yields comparable to those obtained by using specially optimised DNA sequences \cite{Ke2012}, and in the light of this, we have selected a random set of sequences for the patches in the target structure, but such that patches that are adjacent (\textit{i.e.}~bonded) in the correctly-assembled structure have complementary sequences. Like the `protector bricks' in experiment \cite{Ke2012}, unbonded patches at the structure boundaries are given a sequence of 8 consecutive thymines to minimise the chance of their misbonding. Of the 998 particles simulated, 24 have only one interaction with the remaining structure and are unlikely to form stable bonds.

\begin{figure}[tbp]
\centering
\includegraphics{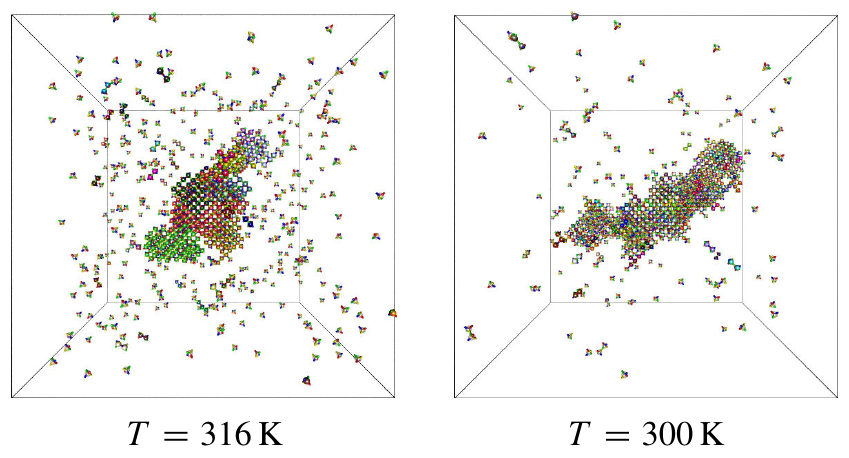}
\caption{Two lower-temperature simulation snapshots demonstrate the formation of kinetic aggregates.}\label{fig-bricks-snapshots-assemblyLowT}
\end{figure}

\begin{figure}[tbp]
\centering
\includegraphics{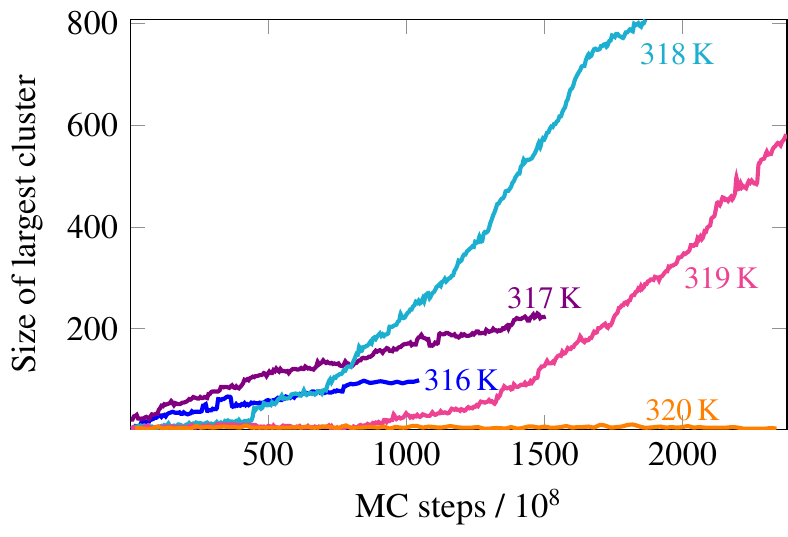}
\caption{The size of the largest correctly bonded cluster is shown as a function of time for several temperatures for a particular set of trajectories.}\label{fig-bricks-clusterSize-vs-time}
\end{figure}

Having determined the appropriate patch sequences, we run simulations at several temperatures, starting from a gas of 998 monomers corresponding to a single target structure \footnote{The salt concentrations used were $\text{\ce{[Na^+]}}=\SI{1}{\mole\per\deci\metre\cubed}$ and $\text{\ce{[Mg^{2+}]}}=\SI{0.08}{\mole\per\deci\metre\cubed}$, which are slightly different from the experimental setup \cite{Ke2012}, but are in the range where the salt concentration dependence formula given by Koehler and Peyret \cite{Koehler2005} is applicable. In the simulations reported here, the simulation box volume was $(62a)^3$; this volume is important when mapping the results to experiment and can significantly affect the nucleation rate, but as we are not directly comparing to a specific experiment, it is an `arbitrary' parameter at this stage.}. At high temperatures, any clusters that form are transient and small. At temperatures around \SI{320}{\kelvin}, however, we observe very interesting behaviour. Several configurations along a particular trajectory at \SI{318}{\kelvin} are shown in Fig.~\ref{fig-bricks-snapshots-assemblyT318}. It is clear from this figure that the system assembles into the designed structure at this temperature. Moreover, several other clusters (which we define as comprising particles each connected to other particles in the cluster by at least one bond corresponding to the designed structure) do grow in addition to the largest one, sometimes connected to the largest one and sometimes not, but at this temperature, they are not sufficiently stable to persist and only one cluster grows at the expense of all others. At long times, the final size of the correctly bonded cluster is approximately 920, corresponding well to the equivalent cluster obtained by relaxing a perfectly assembled structure at this temperature.

If we decrease the simulation temperature (Fig.~\ref{fig-bricks-snapshots-assemblyLowT}), we find ever larger aggregates of incorrectly bonded clusters, \textit{i.e.}~clusters in which patch bonding is not perfectly complementary. The time evolution of the largest cluster size for a particular set of simulations at different temperatures is shown in Fig.~\ref{fig-bricks-clusterSize-vs-time}; we see that at high temperatures, no clusters form; at intermediate temperatures, clusters can grow to large sizes; and at low temperatures, the largest cluster does not grow considerably after an initial growth stage, as other clusters have formed and misbonded, and these `incorrect' bonds do not readily dissociate. At temperatures just below the successful assembly regime, the largest cluster can grow to appreciable sizes, but multiple large correctly bonded clusters typically form, and these then struggle to meet in the correct way, yielding a misformed structure. If we run simulations starting from the fully formed structure, it remains mainly intact to temperatures between \SI{325}{\kelvin} and \SI{330}{\kelvin}; \textit{i.e.}~there is some hysteresis in the transformation.

\begin{figure}[tbp]
\centering
\includegraphics{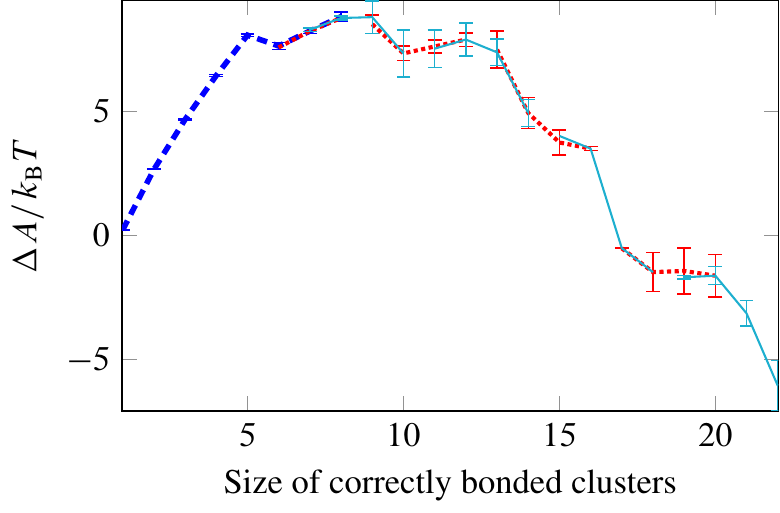}
\caption{The free-energy profile for cluster growth at $T=\SI{319.5}{\kelvin}$. Simulation results from different umbrella sampling windows are depicted in alternating styles to show their overlap. The thick dashed line corresponds to brute-force simulations.}\label{fig-bricks-freeenergybarrier}
\end{figure}

We simulated three additional independent repeats of the simulations discussed above at equidistant temperatures between \SI{310}{\kelvin} and \SI{325}{\kelvin}, and a further 10 runs each at 317, 318 and \SI{319}{\kelvin}, and observed the same qualitative behaviour. The correct structure forms at temperatures between about \SI{317}{\kelvin} and \SI{319}{\kelvin}, but with various lead times before significant growth takes place \footnote{The number of simulations in which the correct structure was found to form out of a total of 14 simulations each was 12 (\SI{319}{\kelvin}), 10 (\SI{318}{\kelvin}) and 4 (\SI{317}{\kelvin}).}. This suggests that there is a free-energy barrier to nucleation that increases with the temperature; the higher the temperature, the rarer the nucleation event, but, by contrast, the smaller the chance of incorrect assembly. To quantify the magnitude of this free-energy barrier, we ran umbrella sampling simulations at \SI{319.5}{\kelvin}, where the free-energy barrier is expected to be relatively small. We plot the free energy as a function of the cluster size in Fig.~\ref{fig-bricks-freeenergybarrier} \footnote{We define the free energy as $\upDelta A(n)/k_\text{B}T \equiv -\ln (N_n/N) \propto -\ln (N_n/V)$, where $N_n$ is the number of clusters of size $n$ and $N$ is the total number of particles. For small $n$, this can be estimated directly from brute-force simulations of the initial state of the system. For larger $n$, $N_n/N$ becomes small and, moreover, rapidly decreases with $n$.  It therefore approaches the probability that the largest cluster in the system is of size $n$.  Beyond $n=7$, we assume that these probabilities are the same, and we calculate free energies using umbrella sampling with the largest cluster size serving as the order parameter.}. The number of clusters of size $n$ per unit volume decreases rapidly with $n$, and in order for the largest cluster in the system to grow beyond just a few particles, a free-energy barrier must be overcome. The critical cluster size at \SI{319.5}{\kelvin} is approximately 10; beyond this size, the free energy predominantly decreases as the cluster grows. However, this decrease is not monotonic, reflecting the fact that certain clusters, typically involving `caged' structures with few dangling particles, are favoured over others; this is not dissimilar to the multi-peaked nucleation barriers seen in Ising-type models \cite{Schneidman2003}. Nevertheless, the principal free-energy barrier  to nucleation at this temperature appears to be relatively small, consistent with the fact that spontaneous growth is (eventually) observed in brute-force simulations. Moreover, brute-force simulations starting from the critical cluster size as determined by umbrella sampling simulations confirm that the critical cluster size has been correctly identified, although the precise structure of a given cluster has a significant effect when considering its propensity to grow or shrink: the cluster size alone is not an optimal order parameter. At lower temperatures, where the nucleation free-energy barrier is very small, several nuclei can form simultaneously and form aggregates. To achieve successful self-assembly, nucleation barriers should be sufficiently high to suppress such in-growth aggregation.

We also performed several additional brute-force simulations with a different random choice of patch sequences; the same overall behaviour is observed, although the precise temperature range at which nucleation occurs varies by a few degrees. Nevertheless, it appears that regardless of the choice of sequence, a slow annealing process from high temperatures will result in the growth of the designed structure, as the system will always pass through the optimal growth regime on cooling. The self-assembly of multiple copies of the target structure in the same simulation box is also successful in roughly the same temperature range.

It is intriguing that the designed structures nucleate reproducibly; however, it is worth looking at the limits of the model and the effects we have neglected. Firstly, some of the most competitive alternative structures are likely to be ones that form with exactly the correct sequence pairing, but with different replicas of the molecules incorporated into the same final structure,  disrupting the geometry of the growing cluster and leading to frustration. On a lattice, this becomes less probable because the system geometry is essentially externally imposed. Moreover, notwithstanding this competition effect, simulating the growth of a single target structure is unlikely to result in bulk assembly statistics \cite{Ouldridge2012}. Secondly, the `patchy' nature of the potential has several implications. Single-stranded DNA particles have a reduced entropy relative to the experimental system because we fix the tetrahedral geometry in advance, likely leading to a relative destabilisation of the single-stranded state, meaning that any melting points we obtain are expected to be higher than in experiment. Moreover, bond angles do not change when interactions involve fewer than 8 base pairs, and we do not consider any hybridisation between parts of domains (\textit{e.g.}~a strand may preferentially bond with parts of domains 1 and 2, but we only consider bonding with either domain 1 or 2). Within our model, there may be several bonding patterns with fewer than 8 matching base pairs of similar strength possible between a pair of patches, which could stabilise some weak bonds entropically, but we do not account for this. Finally, although the 48-nucleotide-long `boundary bricks' seem to be important in experiment, and are likely to be more important for structures more intricate than cubes, we have not simulated them. However, whilst it is certainly important to be aware of these simplifications and omissions in our simulations, the basic physics of self-assembly appears to be captured by our model, and our simulations support the suggestion of Ke~\textit{et~al.}~that initial structure growth is a slow process.

In summary, we have performed lattice MC simulations of a model system designed to mimic the behaviour of DNA bricks studied experimentally by Ke~\textit{et~al.}~\cite{Ke2012}. We have demonstrated that there is a sweet spot in temperature for which the self-assembly of the target structure is successful. Above this temperature range, the monomer phase is entropically favoured, while below it, non-specific bonding results in the growth of large aggregate structures. In experiment, structures are formed via a slow annealing process, passing through the optimum temperature range, and so the desired structures form in reasonable yields. Our simulations support the basic premise that slow nucleation is followed by faster structure growth, as posited by Ke~\textit{et~al.}~\cite{Ke2012}. Finally, the very fact that we use a highly simplified model implies that our results should carry over to other systems, such as (nano)colloids designed with the same properties as DNA bricks. This observation is extremely encouraging, because it suggests that it should be possible to assemble systems consisting of materials other than DNA into complex target structures. This might offer a route to realising the complex colloidal structures proposed in Ref.~\citenum{Halverson2013}.

\begin{acknowledgments}
This work was supported by the European Research Council [Advanced Grant 227758] and the Engineering and Physical Sciences Research Council [Programme Grant EP/I001352/1]. We thank Thomas Ouldridge and Peng Yin for useful discussions.
\end{acknowledgments}

\end{document}